\documentclass[12pt]{article}
\usepackage{graphicx}
\usepackage{amsmath,latexsym,epsfig,epsf,rotate,amssymb}
\newcommand{\be}{\begin{equation}}
\newcommand{\ee}{\end{equation}}
\begin{document}

\title{Test of FSR in the process $e^+e^-\to\pi^+\pi^-\gamma$ at DA$\Phi$NE and extraction of the pion form factor at threshold.
}

\author{G.~Pancheri$^{1)\footnote{e-mail: Giulia.Pancheri@lnf.infn.it}}$
, O.~Shekhovtsova$^{1),2)\footnote{e-mail: shekhovtsova@kipt.kharkov.ua}}$,
 G.~Venanzoni$^{1)\footnote{e-mail: Graziano.Venanzoni@lnf.infn.it}}$ 
\\
\\
\emph{$^{1)}$INFN Laboratori Nazionale di Frascati, Frascati (RM) 00044, Italy} \\
\emph{$^{2)}$NSC ``Kharkov Institute for Physics and Technology'',} \\
\emph{Institute for Theoretical Physics, Kharkov 61108, Ukraine }
} 
\date{}
\maketitle

\begin{abstract}

Effects due to non-pointlike behaviour of pions
in the process  $e^+e^-\to\pi^+\pi^-\gamma$  can arise for hard photons in the final state.
By means of a Monte Carlo event generator, which also includes 
the contribution of the direct decay
$\phi\to\pi^+\pi^-\gamma$,
we estimate these effects in the framework of Resonance Perturbation Theory.
We consider angular cuts used in the
KLOE analysis of the pion form factor at threshold.
A method to reveal the effects of non-pointlike behaviour of pions  in a model-independent way is proposed.

\end{abstract}

\section {Introduction} 
Final state radiation (FSR) is an irreducible background 
to the measurement of the hadronic cross section with
initial state radiation (ISR) events \cite{binner,kloe}.
Differently from ISR, whose accuracy is limited by the 
numerical precision on the evaluation 
of high order QED diagrams (see, for example,~\cite{jadach,phokhara} and discussion there), 
the FSR evaluation relies on specific models for the coupling of  hadrons to photons.
Usually the FSR amplitude in the process $e^+e^-\to\pi^+\pi^-\gamma$
is evaluated in scalar QED (sQED), 
where
the 
pions  are treated as point-like particles and 
the total FSR amplitude is multiplied by the pion form factor 
computed in the VMD model \cite{phokhara,sqed}.
While this assumption is generally valid for relatively 
soft photons, 
it can fail for low values of the invariant mass of the hadronic system, 
{\it i.e.} when the intermediate hadrons are far off shell.
In this case possible extensions for FSR, beyond sQED, can be 
considered.

As we will show in the following, 
the most general form for FSR consistent with gauge invariance, 
charge conjugation symmetry and photon crossing symmetry
can be expressed in term of three form factors, $f_i$, 
each depending on three independent variables~\cite{our,fsr1}.
While such a decomposition is general, $f_i$ are model dependent.
In the paper~\cite{our} the prediction for  $f_i$ in the framework of
 the Resonance Perturbation Theory (RPT) was considered.
RPT is a model based on
Chiral Perturbation  Theory ($\chi$PT)
with the explicit inclusion of the vector  
and  axial--vector mesons, $\rho_0(770)$ and  $a_1(1260)$~\cite{Ecker_89}.
Whereas $\chi$PT gives correct predictions on the pion form factor at
very low energy, RPT is the appropriate framework to describe the pion form
factor, and to satisfy QCD high energy behaviour, at intermediate energies
($E \sim m_\rho$)~\cite{Ecker_89}~\footnote{In that paper
it was shown that the coupling constants of the effective chiral
lagrangian at the order $p^4$ are essentially saturated by meson
resonance exchange.}.


We point out that the FSR process,
besides being of interest as an 
important background source, is of interest by itself, since it allows to  
get information about pion-photon interaction at low energies.

In this paper we will present the results obtained by
implementing the RPT amplitude for FSR 
 into the Monte Carlo event generator 
for the process $e^+e^-\to\pi^+\pi^-\gamma$~\cite{binner}. We also included the rare
decay $\phi\to\pi^+\pi^-\gamma$~\cite{graz} in our generator.

Since 
 most of the effects arising 
in the FSR are model-dependent, 
we conclude this paper by suggesting a way, based on Monte Carlo,
 to test possible effects 
beyond sQED, in a model-independent way.

\section{ General parametrization for $\pi^+\pi^-\gamma$ final state} 
The cross section of the reaction
$$e^+(p_1)+e^-(p_2)\to\pi^+(p_+)\pi^-(p_-)\gamma(k) $$ 
with the photon emitted in the final state,
can be written as
\begin{equation}
d\sigma=\frac{1}{2s(2\pi)^5}\int
\delta^4(P-p_+-p_--k)\frac{d^3p_+d^3p_-d^3k}{8E_+E_-\omega}|M|^2 ,
\end{equation}
where $P=p_1+p_2$, $s=P^2$ and
\be\label{fsr_ampl}
M=\frac{e}{s}M^{\mu\nu}\bar u(-p_1)\gamma_\mu u(p_2)\epsilon^\star_{\nu} .
\ee

The tensor $M_F^{\mu\nu}$ describing the process
$\gamma^*(P)\to\pi^+(p_+)\pi^-(p_-)\gamma(k)$ is model-dependent.
However, based on charge-conjugation symmetry,
photon crossing symmetry and gauge invariance it
can be expressed by three gauge invariant tensors
(see Appendix A in Ref.~\cite{our}):
\begin{eqnarray}\label{eqn:fsr}
&&M^{\mu \nu }(P,k,l)=-ie^{2}(\tau _{1}^{\mu \nu }f_{1}+\tau
_{2}^{\mu \nu }f_{2}+\tau _{3}^{\mu \nu }f_{3})\equiv
-ie^{2}M_{F}^{\mu \nu }(P,k,l),  \; \; \; \; l=p_+-p_-, \label{fsr_str}\\
&&\tau _{1}^{\mu \nu }=k^{\mu }P^{\nu }-g^{\mu \nu }k\cdot P,  \nonumber \\
&&\tau _{2}^{\mu \nu }=k\cdot l(l^{\mu }P^{\nu }-g^{\mu \nu
}k\cdot
l)+l^{\nu }(k^{\mu }k \cdot l-l^{\mu }k \cdot P),  \nonumber \\
&&\tau _{3}^{\mu \nu }=P^{2}(g^{\mu \nu }k\cdot l-k^{\mu }l^{\nu
})+P^{\mu }(l^{\nu }k\cdot P-P^{\nu }k\cdot l). \nonumber
\end{eqnarray}

While the decomposition (\ref{eqn:fsr}) is general,
the exact value of the scalar functions $f_{i}$ (form factors),
each depending in terms of three independent variables, 
are determined by the specific FSR models. 

In sQED for the functions $f_i$ we have \cite{sqed}
\begin{equation}
f_1^{sQED}=\frac{2k\cdot P}{(k\cdot P)^2-(k\cdot l)^2}, \; \; \;
\; f_2^{sQED}=\frac{-2}{(k\cdot P)^2-(k\cdot l)^2}, \;\; \; \;
f_3^{sQED}=0,
\end{equation}
Because of Low's theorem, these equations imply that for $k\to 0$  we have
\begin{equation}\label{funct_0}
\mathrm{lim}_{k\to 0}f_1=\frac{2k\cdot P F_\pi(P^2)}{(k\cdot
P)^2-(k\cdot l)^2}, \hspace{1.5em} \mathrm{lim}_{k\to
0}f_2=\frac{-2 F_\pi(P^2)}{(k\cdot P)^2-(k\cdot l)^2},
\hspace{1.5em} \mathrm{lim}_{k\to 0}f_3=0,
\end{equation}
where $F_\pi$ is the pion form factor describing the interaction $\gamma^*\to\pi^+\pi^-$. 
Thus, for soft photon radiation, the FSR tensor is  expressed in term of one form factor $F_\pi(P^2)$, but in general we have  three independent form factors describing the FSR process.

It is convenient to rewrite the form factors $f_i$ as
\begin{equation}\label{f}
f_{i}=f_{i}^{sQED}+\Delta f_{i},
\end{equation}
where the functions $\Delta f_i$ are the contributions to the form factors beyond sQED, and are determined by FSR model.

\subsection{RPT contribution to FSR}

The functions $\Delta f_i$
 have been calculated in the framework of RPT in \cite{our}:
\begin{eqnarray}\label{d_f}
\Delta f_{1} &=&\frac{F_{V}^{2}-2F_{V}G_{V}}{f_{\pi }^{2}}\biggl(\frac{1}{%
m_{\rho }^{2}}+\frac{1}{m_{\rho }^{2}- P^2-\mathrm{i}m_\rho\Gamma_\rho(P^2)}\biggr)  \nonumber \\
&-&\frac{F_{A}^{2}}{f_{\pi }^{2}m_{a}^{2}}\biggl[ 2+\frac{(k\cdot l)^{2}}{%
D(l)D(-l)}+\frac{(P^{2}+k\cdot P)[4m_{a}^{2}-(P^{2}+l^{2}+2k\cdot P)] }{
8D(l)D(-l)}\biggr],  \label{eq:delta-f1} \\
\Delta f_{2} &=&-\frac{F_{A}^{2}}{f_{\pi }^{2}m_{a}^{2}}\frac{%
4m_{a}^{2}-(P^{2}+l^{2}+2k\cdot P)}{8D(l)D(-l)} ,  \label{eq:delta-f2} \\
\Delta f_{3} &=&\frac{F_{A}^{2}}{f_{\pi }^{2}m_{a}^{2}}\frac{k\cdot l}{%
2D(l)D(-l)} , \; \; \; D(l)=m_a^2-(Q^2+l^2+2kQ+4kl)/4 .
\end{eqnarray}

For notations and details of the calculation  we refer a reader 
to~\cite{our}.
$F_V$, $G_V$ and $F_A$ are parameters of the model. 
The parameters $F_V$, $G_V$ as well as $m_\rho$ have been estimated by a fit to the pion form factor from 
$e^+e^-\to\pi^+\pi^-$ data 
(see next Section).  For $a_1$ meson  we take $m_a=1.23$ GeV and $F_A=0.122$ GeV that corresponds to the mean value of the experimental decay width $\Gamma(a_1\to\pi\gamma)=640\pm 246$ keV \cite{pdg}.

\subsection{Pion form factor in RPT} 

The pion form factor, that describes $\rho-\omega$ mixing and includes also the first excited $\rho'$-meson state, can be written  as:
\begin{equation}\label{formfact}
F_\pi(q^2)=1+\frac{F_V G_V}{f_\pi^2}B_\rho(q^2)
\Biggl(1-\frac{\Pi_{\rho\omega}}{3q^2}B_\omega(q^2)\Biggr)+
\frac{F_{V1} G_{V1}}{f_\pi^2}B_{\rho'}(q^2) ,
\end{equation}
where 
\begin{equation}\label{rho}
B_{r}(q^2)=\frac{q^2}{m_r^2-q^2-\mathrm i m_r
\Gamma_r(q^2)} ,
\end{equation}
$q^2$ is the virtuality of the photon, $f_\pi=92.4$ MeV  and the value $\Pi_{\rho\omega}$ describes $\rho$-$\omega$ mixing (see below). More detailed description of the pion form factor can be found 
elsewhere~\cite{connell,our_fut}.

An energy
dependent width  is considered for the  $\rho$ and $\rho'$ mesons:
\begin{equation}\label{gamma}
\Gamma_\rho(q^2)=\Gamma_\rho\sqrt{\frac{m_\rho^2}{q^2}}
\Biggl(\frac{q^2-4m_\pi^2}{m_\rho^2-4m_\pi^2}\Biggr)^{3/2}\cdot \Theta(q^2-4m_\pi^2),
\end{equation} while for 
the $\omega$--meson a constant width is used,
$\Gamma_\omega=8.68$ MeV, and $m_\omega=782.7$ MeV. 
We assume the parameter $\Pi_{\rho\omega}$, 
that determines $\rho$-$\omega$ mixing, is a constant and we relate it to 
the branching fraction $Br(\omega\to\pi^+\pi^-)$:
\begin{equation}\label{br_om}
Br(\omega\to\pi^+\pi^-)=\displaystyle\frac{\mathstrut
|\Pi_{\rho\omega}|^2}{\Gamma_\rho \Gamma_\omega m_\rho^2} .
\end{equation}

The fit of our parametrization to the pion form factor 
from the CMD-2 data~\cite{cmd2} gives~\cite{our_fut}:
\begin{eqnarray*}\label{num_with}
&& m_\rho=774.97\pm 1.4 \text{ MeV} , \; \; \; \Pi_{\rho\omega}=-2774\pm 291.2 \text{ MeV}^2 ,
\nonumber \\&& \Gamma_\rho=145.21\pm2.6 \text{ MeV} , \; \; \; F_V=154.22\pm0.5 \text{ MeV}
\end{eqnarray*}
and $$m_\rho'=1.2\pm 0.2 \text{ GeV} , \; \; \;  \Gamma_{\rho'}=400\pm160 \text{ MeV} , \; \; \;
F_{V1}=13.19\pm18.59 \text{ MeV} ,$$ $\chi^2/\nu=
0.853$.
Then  $G_V=64.6\pm0.3$ MeV and
$Br(\omega\to\pi^+\pi^-)=(0.96\pm0.19) \%$.
Expanding the pion form factor in the region $s<0.35$ GeV$^2$
as in~\cite{eid_jeg}:
\be\label{decomp_fpi}
F_\pi(q^2)\simeq 1+
p_1\cdot q^2+p_2\cdot q^4 ,
\ee
gives the following values: $p_1=1.15\pm 0.06$ GeV$^{-2}$,  $p_2=9.06\pm 0.25$ GeV$^{-4}$, $\chi^2/\nu\simeq 0.13$.

\subsection {Scalar contribution to FSR}

At $s=m_\phi^2$ an additional contribution to the final state $\pi^+\pi^-\gamma$ is given by 
the direct rare decay $\phi\to \pi^+\pi^-\gamma$. As it was shown in~\cite{graz,gino} this process affects the form factor $f_1$ of Eq.({\ref{eqn:fsr}}):
\be\label{phi_ampl}
M_\phi=\frac{ie^3}{s}F_\phi(Q^2)\bar u(-p_1)\gamma_\mu u(p_2)\epsilon^*_\nu\tau_1^{\mu\nu} ,
\ee
$$F_\phi(Q^2)=\frac{g_{\phi\gamma}f_\phi}{s-m_\phi^2+im_\phi\Gamma_\phi} , \; \; Q=p_++p_- .$$

The $\phi$ direct decay is assumed to proceed through the $f_0$ intermediate state:
 $\phi\to f_0\gamma\to\pi^+\pi^-\gamma$,  and its mechanism is 
described by a single form factor $f_\phi(Q^2)$. 

We consider the model described in~\cite{ach_sol} where the 
$\phi \to f_0 \gamma$ decay amplitude 
is generated dynamically
through the loop of charged kaons.
The form factor $f_\phi$  reads:
\be
f_\phi^{K^+ K^-}(Q^2) = 
\frac {g_{\phi K^+ K^-} g_{f_0 \pi^+ \pi^-} g_{f_0 K^+ K^-}}
{2\pi^2 m^2_{K} (m_{f_0}^2 -Q^2 +Re\Pi_{f_0}(m_{f_0}^2) - 
\Pi_{f_0}(Q^2))}
I\left ( \frac {m_\phi^2}{m_K^2},\frac {Q^2}{m_K^2} \right )
e^{i\delta_B(Q^2)} ,
\ee
where 
$I(.,.)$ is a function known in analytic
form~\cite{graz,close} and $\delta_B(Q^2) = b\sqrt{Q^2-4 m_{\pi}^2} $, $b=75^o/$GeV~\cite{achasov}. The term $Re\Pi_{f_0}(m_{f_0}^2) - \Pi_{f_0}(Q^2)$ takes into account the finite width corrections to the $f_0$ propagator~\cite{ach_sol}.
A fit to the KLOE data $\phi\to\pi^0\pi^0\gamma$~\footnote{$\Gamma(f_0\to\pi^+\pi^-)= \frac{2}{3}
\Gamma(f_0\to\pi\pi)$}  gives
the following values of the parameters~\cite{kloepi0}:
\be
m_{f_0}= 0.962~{\rm GeV},
\,\, \frac{g^2_{f_0 K^+ K^-}}{4\pi} = 1.29~{\rm GeV}^2, 
\,\, \frac{g^2_{f_0 K^+ K^-}}{g^2_{f_0 \pi^+ \pi^-}} = 3.22.
\ee

A refined version of this model 
includes the presence of  the $\sigma$ meson in the intermediate 
state~\cite{achasov,kloepi0,czyz}. 
Such an extension of the model improves the description of the data at low $Q^2$ and will be considered in a forthcoming paper~\cite{our_fut}. Therefore, for the following results, 
we will only consider the presence of the  $f_0$ meson. This will not affect our proposal, to be discussed in Sec. 4.

\subsection {Other contributions}
We included in our program the
channel $\gamma^*\to\rho^\pm\pi^\mp\to\pi^+\pi^-\gamma$,
whose amplitude has been evaluated in RPT model.
However, in agreement with the calculation given in~\cite{our},
we found a negligible contribution of this channel and, for the sake of simplicity,  we discard the effects on the following results.
We also did not include the contribution due to two-resonance intermediate 
states as, for example,  $\phi\to\rho\pi$, which was found to be  negligible in the $\pi^0\pi^0\gamma$ final state \cite{kloepi0}.

\begin{figure}[tbp]
\label{fig1}
\par
\parbox{1.05\textwidth}{
\includegraphics[width=0.5\textwidth,height=0.5\textwidth]
{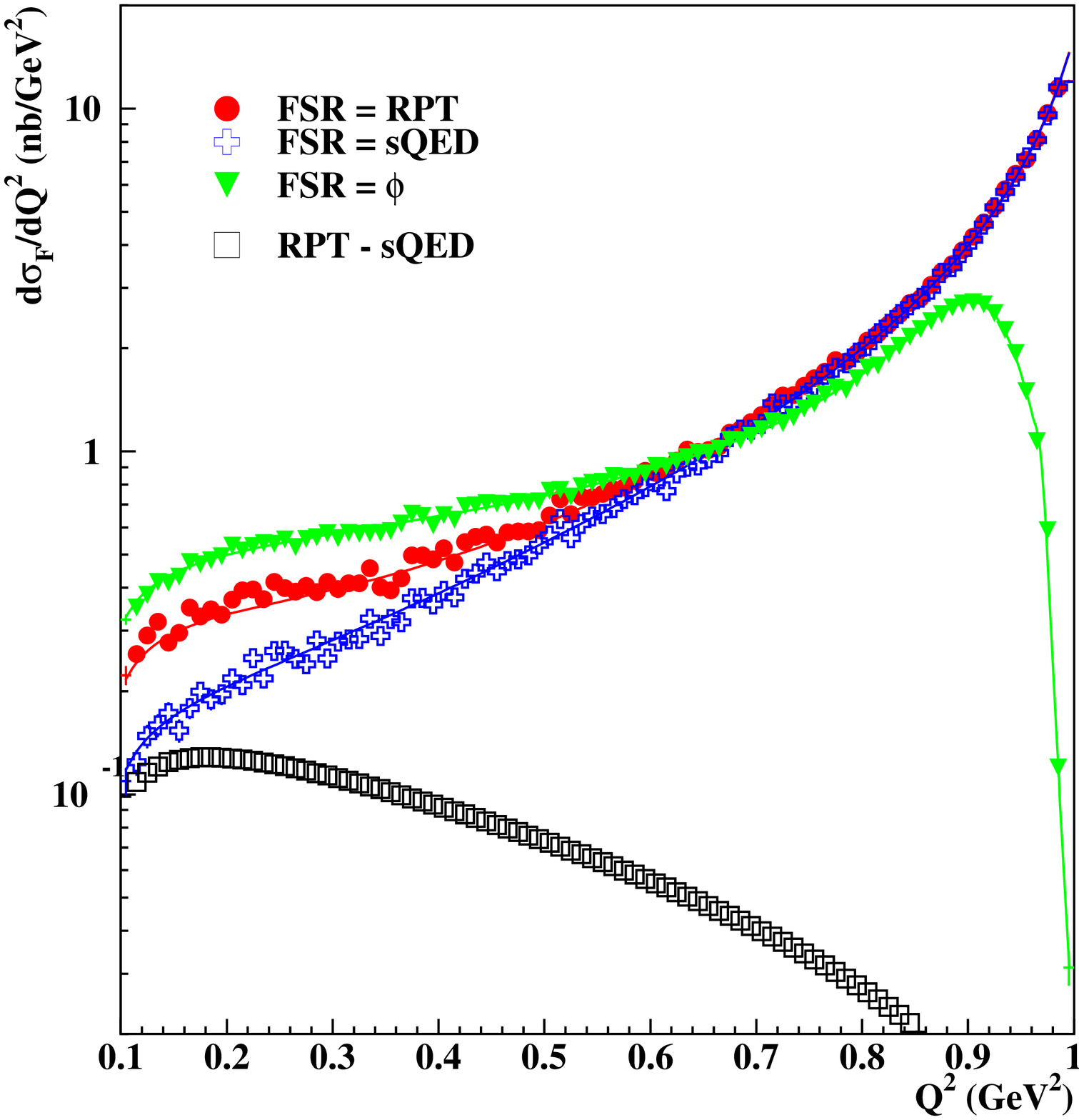}
\hspace{0.5cm}
\includegraphics[width=0.5\textwidth,height=0.5\textwidth]
{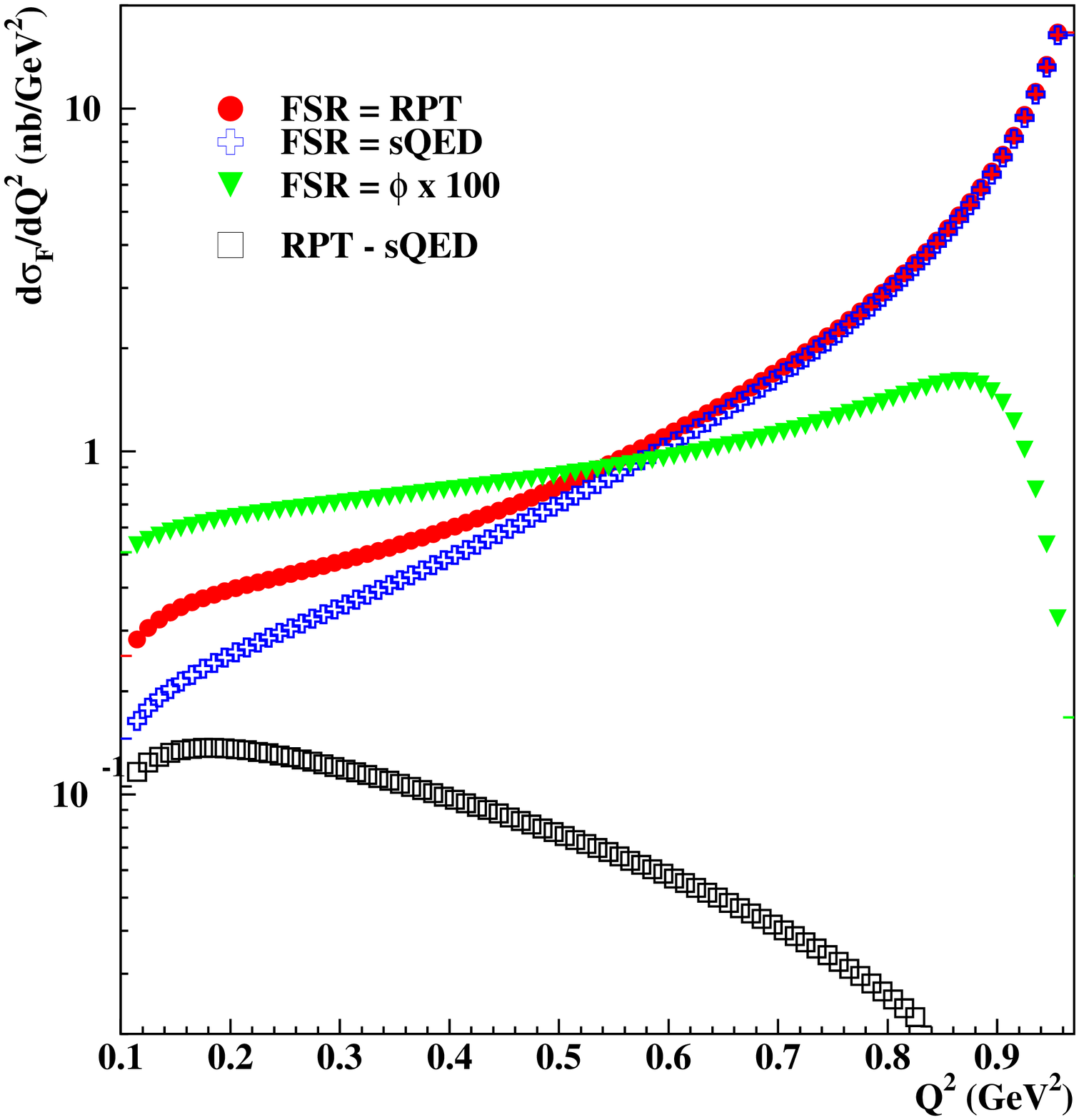}
}
\caption{Contribution to 
the FSR cross section $d\sigma_{F}/dQ^2$ 
in the region  $0^\circ\leq\theta_\gamma\leq 180^\circ$, 
 $0^\circ\leq\theta_\pi\leq 180^\circ$.
 RPT is represented by circles, sQED by crosses, $\phi$ by triangles, 
while the difference between RPT and sQED is indicated by squares.
({\it Left}) corresponds to $s=m_\phi^2$, 
({\it right}) to $s=1$ GeV$^2$ (i.e. below the $\phi$ resonance), 
where the $\phi$ resonant contribution has been amplified by a factor $100$. 
In ({\it left})   
the results from our event generator are compared with the
 analytic calculation, shown by  solid line. 
}
\end{figure}

\section{Numerical results} 

The differential cross section for the reaction 
$e^+e^-\to\pi^+\pi^-\gamma$, where 
the FSR amplitude ($M_{FSR}$) 
receives contributions both from RPT ($M_{RPT}$) 
 and the $\phi\to\pi^+\pi^-\gamma$ decay  ($M_{\phi}$) 
 can be written as:
\begin{eqnarray}\label{cross_sect}
d\sigma_T &\sim& |M_{ISR}+M_{FSR} |^2 = 
d\sigma_{I}+d\sigma_{F}+d\sigma_{IF},\\\nonumber
d\sigma_{I}& \sim & |M_{ISR}|^2, \\\nonumber
d\sigma_{F}&\sim & |M_{RPT}|^2 +
|M_\phi|^2+2\mathrm{Re}\{M_{RPT}\cdot M_\phi^*\} , \\ \nonumber 
d\sigma_{IF}& \sim & 2\mathrm{Re}\{M_{ISR}\cdot (M_{RPT}+ M_\phi)^*\} . \\ \nonumber 
\end{eqnarray}
The interference term $d\sigma_{IF}$ is equal to zero for 
symmetric cuts on the polar angle of the pions~\cite{binner}.

The different contributions to the FSR 
differential cross section $d\sigma_{F}$, evaluated at  $s=m_\phi^2$, 
are shown for in Fig.1, {\it left}, 
 for the full angular range $0^\circ\leq\theta_\gamma\leq 180^\circ , 
0^\circ\leq\theta_\pi\leq 180^\circ$. 
A good agremeent between the results of the  
Monte Carlo simulation (points), with the analytic 
prediction (solid line) is found.
It can be noted that at low $Q^2$ the $\phi$ resonant contribution 
(i.e. the term proportional to $|M_\phi|^2$ in Eq.~(\ref{cross_sect})) 
is quite large
and, therefore, the additional contribution beyond sQED,
can be revealed only in the case of destructive interference between  the two amplitudes ($\mathrm{Re}(M_{RPT}\cdot  M_\phi^*)<0$). 
Published data from  KLOE experiment~\cite{kloe}
are in favour of this assumption, which we will use in the following.
\begin{figure}[tbp]
\label{fig2}
\par
\parbox{1.05\textwidth}{\hspace{-0.4cm}
\includegraphics[width=0.5\textwidth,height=0.5\textwidth]{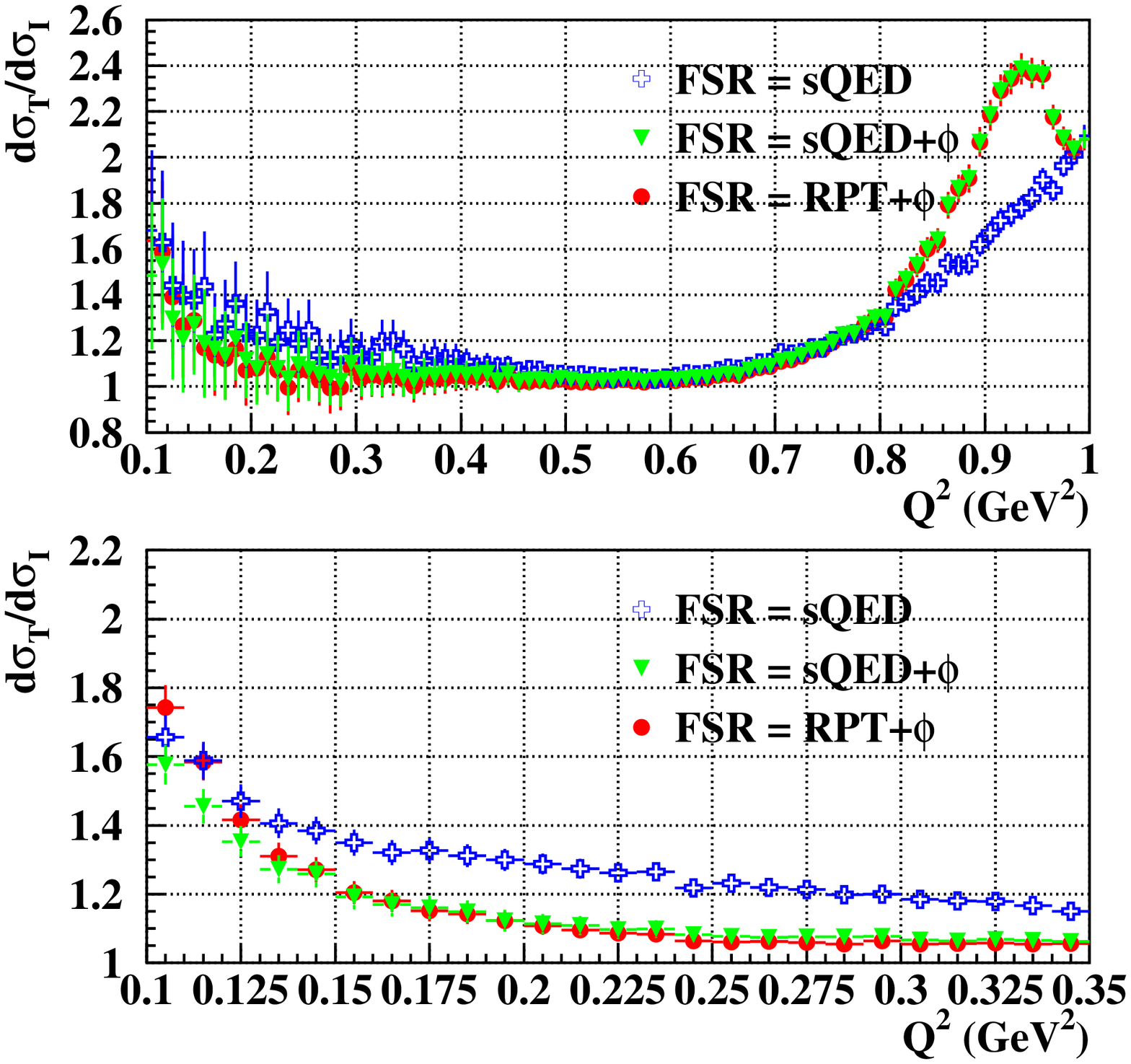}
\hspace{0.2cm}
\includegraphics[width=0.5\textwidth,height=0.5\textwidth]{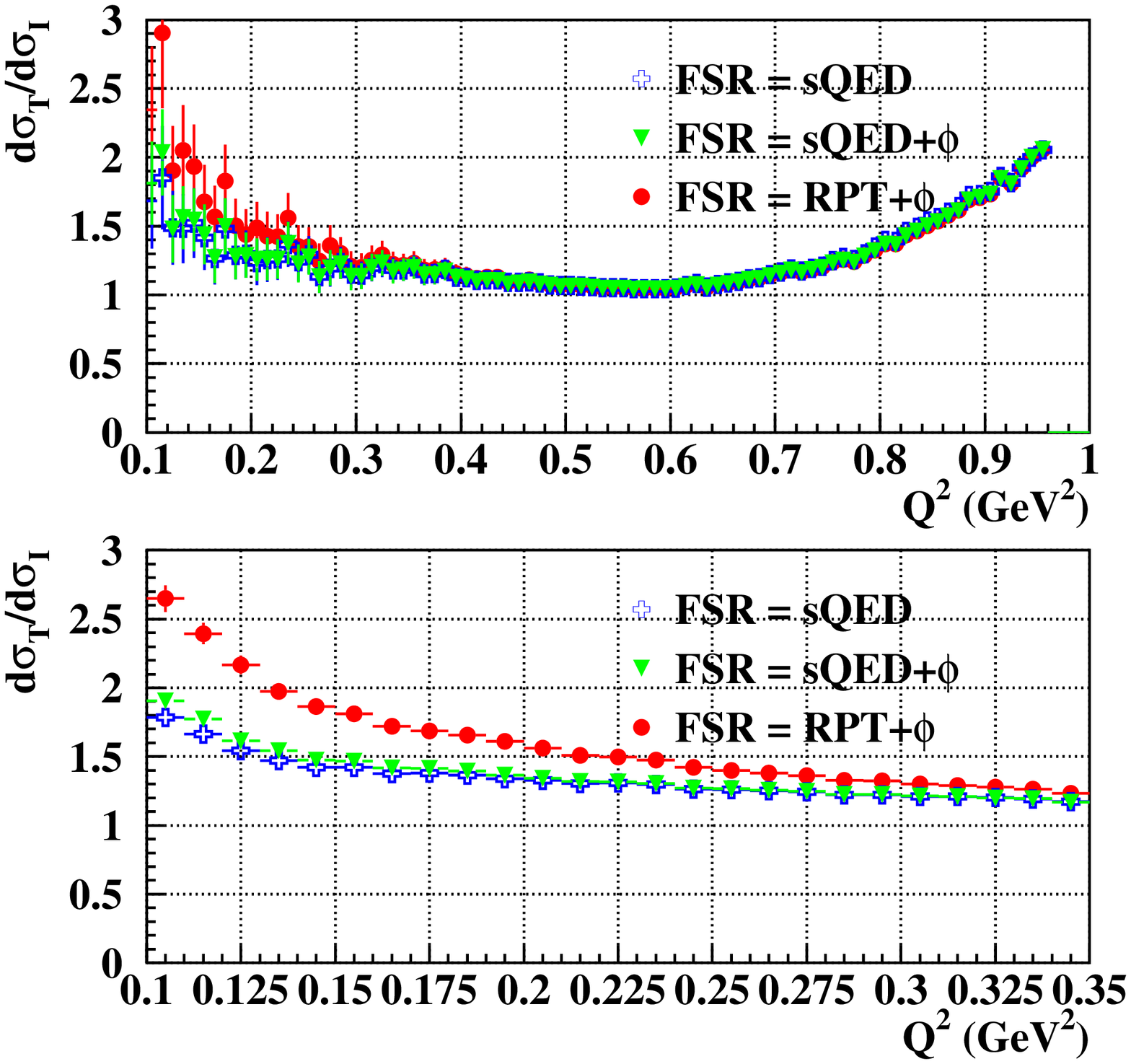}}
\caption{The ratio $d\sigma_T/d\sigma_I$ as function of 
the invariant mass of the two pions, in the region  $50^\circ\leq\theta_\gamma\leq 130^\circ$,
$50^\circ\leq\theta_\pi\leq 130^\circ$, for  different models of FSR.
({\it Left}) refers to $s=m_\phi^2$, ({\it right}) to $s=1$ GeV$^2$.
}
\end{figure}

In Fig.2 we show the values of $d\sigma_T/d\sigma_I$ for the angular 
cuts of the KLOE large angle analysis
$50^\circ\leq\theta_\gamma\leq 130^\circ$, 
$50^\circ\leq\theta_\pi\leq 130^\circ$~\cite{kloe_large} , with  and
 without contributions from 
RPT and $\phi$ direct decay, 
for a  hard photon radiation with energies $E_\gamma>20$ MeV. 
\begin{figure}
\label{fig3}
\par
\parbox{1.05\textwidth}{\hspace{-0.4cm}
\includegraphics[width=0.5\textwidth,height=0.5\textwidth]{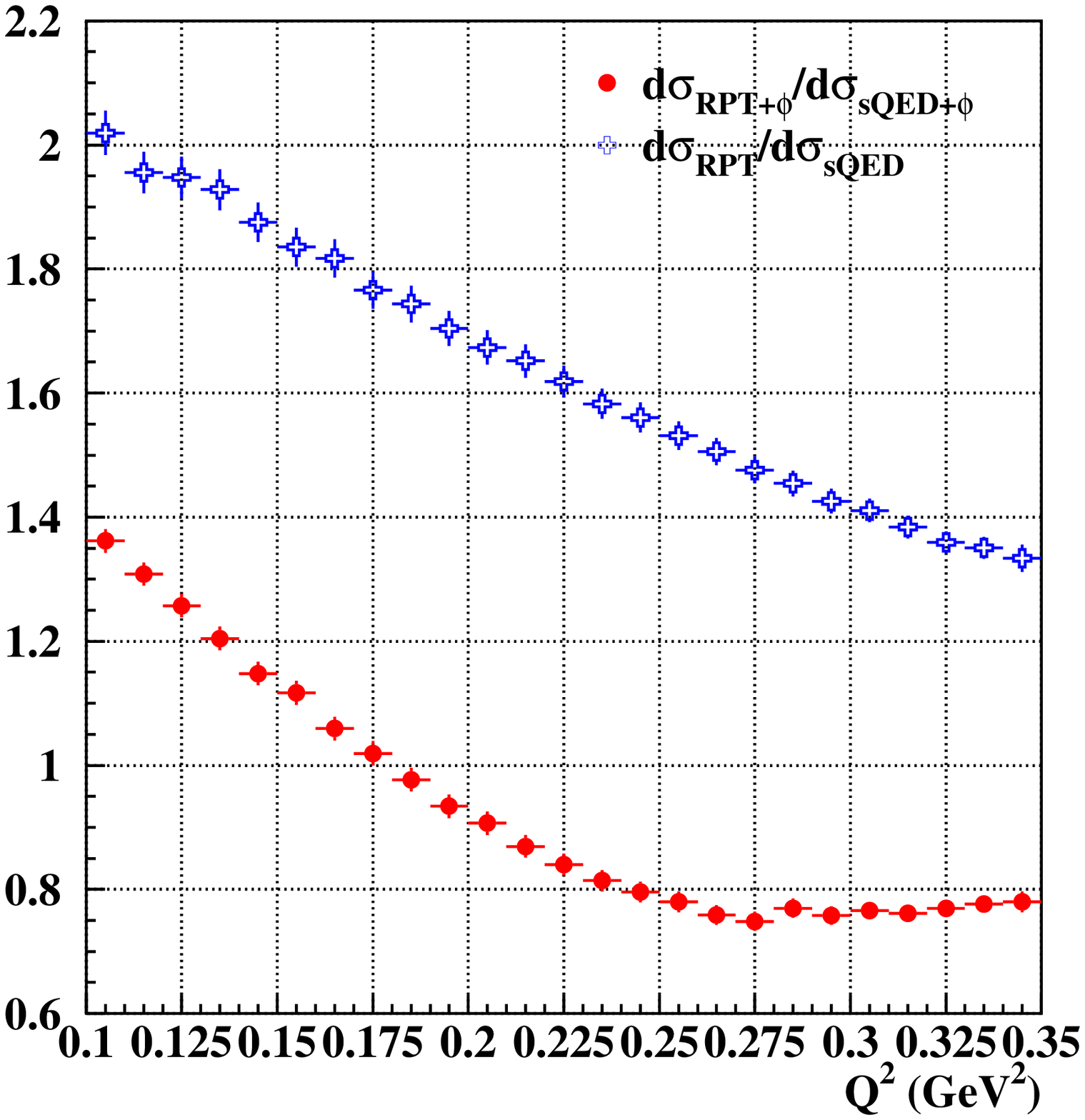}\hspace{0.2cm}
\includegraphics[width=0.5\textwidth,height=0.5\textwidth]{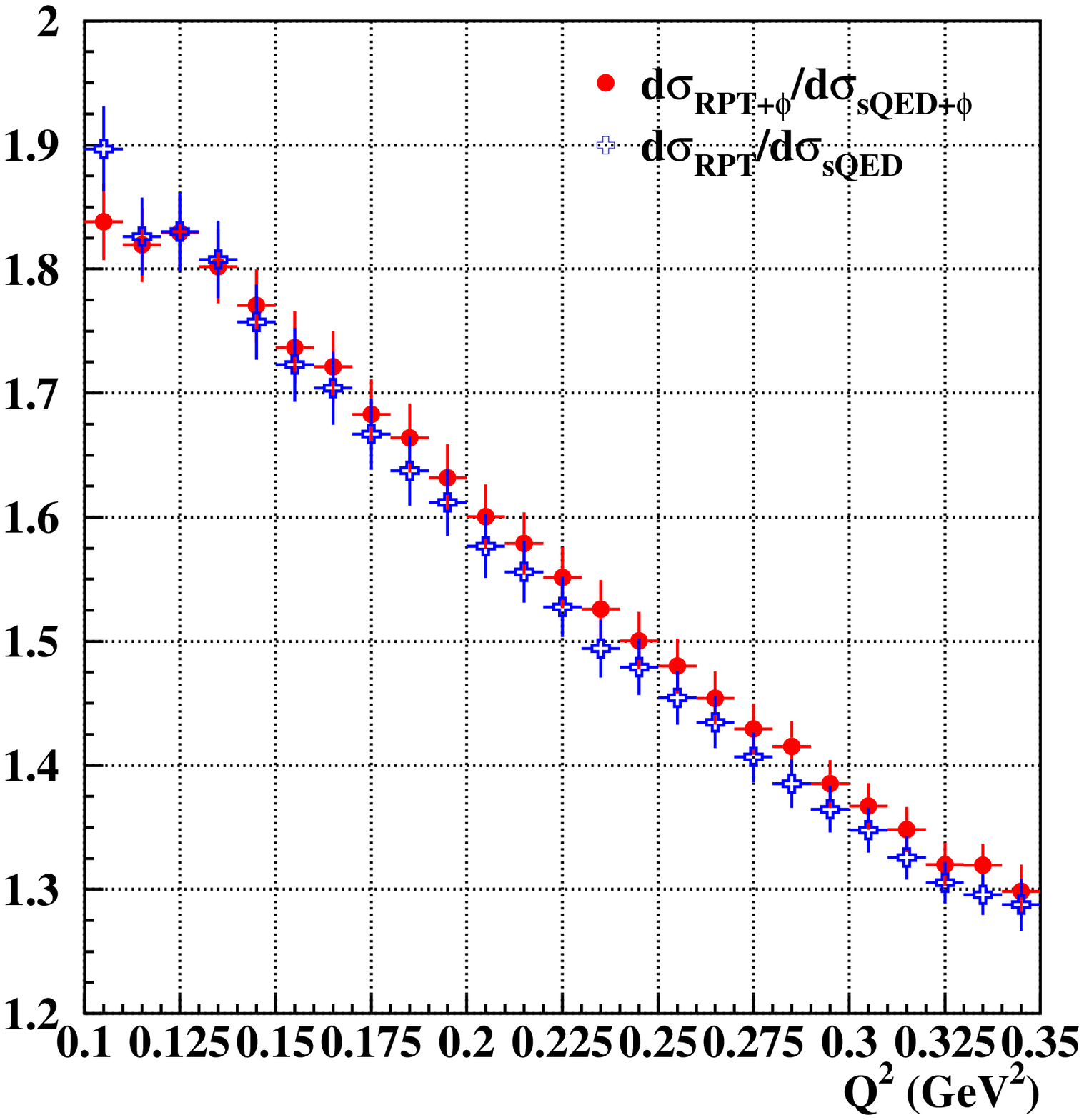}}
\vspace{-0.5cm} 
\caption{
Ratio of FSR cross section in the framework of RPT, respect to sQED, when the $\phi$ contribution is (or not) taken account.  The angular region is
$50^\circ\leq\theta_\gamma\leq 130^\circ$, 
$50^\circ\leq\theta_\pi\leq 130^\circ$.
({\it Left}) is for $s=m_\phi^2$, ({\it right}) is for $s=1$ GeV$^2$.
}
\end{figure}

Three distinctive features can be noted:{\it (1)} the 
peak at about $1$ GeV$^2$ corresponds 
to the $f_0$ intermediate state for the $\phi\to\pi\pi\gamma$ amplitude;
{\it (2)} the presence of RPT terms in the FSR are relevant at low $Q^2$, where
they give
 an additional contribution up to $40\%$ on the ratio  
$d\sigma_{RPT+\phi}/d\sigma_{sQED+\phi}$, (as shown in Fig.~3, {\it left}); {\it (3)} the negative interference with the $\phi$ direct decay amplitude reduces $d\sigma_F$ and its dependence on FSR model at low $Q^2$ (see Fig.~2, {\it left}, down, and  Fig.3, {\it left}).
\begin{figure}
\label{fig4}
\par
\parbox{1.05\textwidth}{\hspace{-0.3cm}
\includegraphics[width=0.5\textwidth,height=0.5\textwidth]{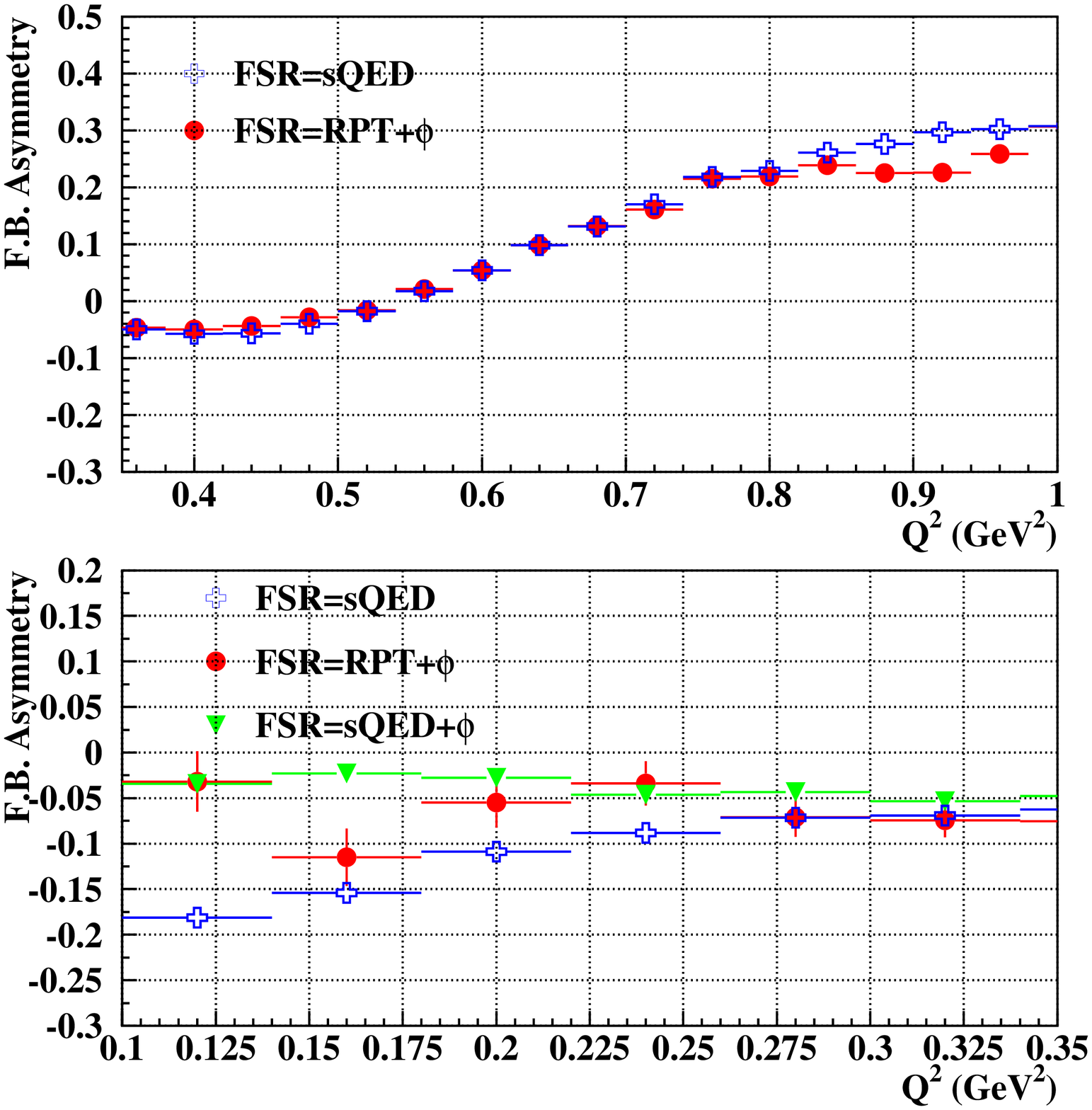}
\hspace{0.2cm}
\includegraphics[width=0.5\textwidth,height=0.5\textwidth]{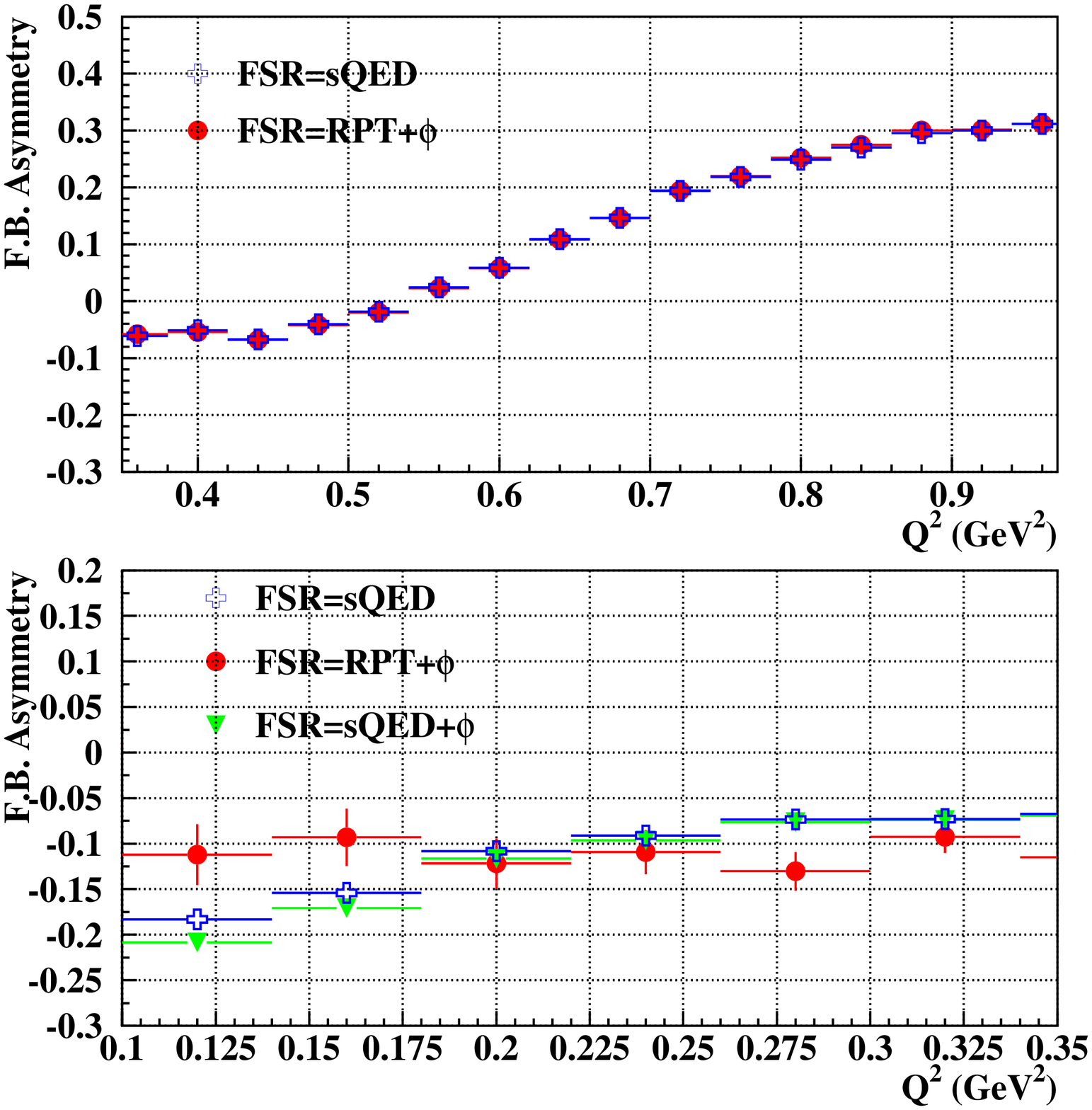}}
\vspace{-0.5cm} 
\caption{
Forward-backward asymmetry, in the kinematical region
$50^\circ\leq\theta_\gamma\leq 130^\circ$, 
$50^\circ\leq\theta_\pi\leq 130^\circ$, 
when $RPT$ and $\phi$ contributions are included, compared with the asymmetry calculated in sQED.
({\it Left}) is for $s=m_\phi^2$, ({\it right}) is for $s=1$ GeV$^2$.
}
\end{figure}

Fig.~4, {\it left}, shows the effects of RPT and $\phi$ terms in the 
forward--backward asymmetry. As expected the presence of RPT gives relevant 
effects at low $Q^2$ 
region, while the presence of a bump at high $Q^2$ is due to the $\phi$ direct decay.

In order to reduce the background from  $\phi\to3\pi$ decay on the measurement of the pion form factor at threshold, 
KLOE has taken more than $200$ pb$^{-1}$ of data at 1 GeV~\cite{stef}.
In this case, as shown in Fig.~1, {\it right}, the 
$\phi$ resonant contribution  is suppressed ($d\sigma_T$ with and without the $\phi$ direct decay almost coincide), see also Figs.2 and 3, {\it right}. 
Therefore the main contribution  beyond sQED to the FSR cross section and the asymmetry (see Fig. 4, {\it right}) comes from RPT. 


\section{Model-independent test of FSR and extraction 
of pion form factor at threshold}
Contributions to FSR beyond sQED, as in the case of RPT,
can lead to sizeable effects on the cross 
section and asymmetry at threshold, as shown in Figs. 3 and 4.
Precise measurement of the pion form factor in this region needs 
to control them at the required level of accuracy.
This looks like a rather difficult task, if one thinks that  effects
 beyond sQED, as well as the contribution from $\phi\to\pi^+\pi^-\gamma$,
are model dependent.

One can think to construct a general amplitude for the 
$e^+e^-\to\pi^+\pi^-\gamma$, according to some underlying theory, and try to determine 
the free parameters
by a constrained fit on specific variables 
(like mass spectrum, charge and forward-backward asymmetry, angular distribution,
etc...). Particularly for the charge asymmetry, it has 
 been proved to be a powerful tool to discriminate between
different models of  $\phi\to\pi^+\pi^-\gamma$~\cite{czyz}.
 However when the number of the parameters is large,
correlations between the parameters of the model 
can arise and spoil the effective power of these fits.
The situation becames even worse if also the pion form factor has to be extracted with the same data. 
As an example, in the case of RPT model,  if we consider only $\rho$ and $\omega$ contribution to the pion form factor and the $\rho$ and $a_1$ contribution to FSR the number of free parameters is already six. The presence of the $\phi$ direct decay adds additional free parameters.

The possibility to determine some of the parameters by external data can strongly help, as in the case of the $\phi\to\pi^+\pi^-\gamma$ amplitude, 
which can be determined by the $\pi^0\pi^0\gamma$ channel 
copiously produced at DA$\Phi$NE.
An additional source of information which will be used to determine the
contributions to FSR beyond sQED in a model-dependent way,
is the energy dependence of the FSR amplitude 
on the $e^+e^-$ invariant mass $s$.

Let us write the differential cross section for the emission of one photon in
 the process $e^+e^-\to\pi^+\pi^-\gamma$ as function of the 
invariant mass of the two pions: 
\begin{equation}\label{spectr_low}  
\Big(\frac{d\sigma_T}{dQ^2}\Big)_s=
|F_\pi(Q^2)|^2 H_s(Q^2)+\Big(\frac{d\sigma_F}{dQ^2}\Big)_s,
\end{equation}
where $H_s(Q^2)$ is the so called radiation function, which accounts for ISR
emission,
and $\Big(\mathstrut\frac{\displaystyle d\sigma_F}{dQ^2}\Big)_s$ is the differential cross section
for the emission of a photon in the final state.
We explicity put in evidence the dependence of each quantity on the 
$e^+e^-$ invariant mass ($s$).
Since we will consider only symmetric angular cuts for pions, 
the  interference term between initial and final state radiation has been neglected.

The FSR differential cross section, $\Big(\mathstrut\frac{\displaystyle d\sigma_F}{dQ^2}\Big)_s$, is dominated at relatively high $Q^2$ by the contribution coming from sQED ($M_{sQED}$) and $\phi$ direct decay ($M_{\phi}$):  
\be
\Big(\frac{d\sigma_{sQED+\phi}}{dQ^2}\Big)_s 
\sim  |M_{sQED}+ M_{\phi} |^2 .
\ee
Contributions beyond sQED ($\Delta M$) are expected to be important at low $Q^2$.
They introduce  an additional term ($\Delta M$) in the above expression:
\begin{eqnarray}
\Big(\frac{d\sigma_F}{dQ^2}\Big)_s  &\sim &| M_{sQED}+ \Delta M + M_{\phi} |^2 = \\
& = & |M_{sQED}+ M_{\phi} |^2 + |\Delta M|^2 + 2\mathrm{Re}\Big\{\Delta M \cdot
(M_{sQED}+M_\phi)^* \Big\} .
\end{eqnarray}


We will now consider the following quantity:
\begin{equation}\label{y}
Y_s(Q^2)=\frac{
\Big (\frac{d\sigma_T}{dQ^2}\Big)_s-
\Big(\frac{d\sigma_{sQED+\phi}}{dQ^2}\Big)_s}{H_s(Q^2)} =  
|F_\pi(Q^2)|^2+\Delta F_s(Q^2),
\end{equation}
where $\Delta F_s \sim \Big(|\Delta M|^2 + 
2\mathrm{Re}\Big\{\Delta M \cdot
(M_{sQED}+M_\phi)^* \Big\}\Big)_s/ H_s$.

If no contribution beyond sQED is present ($\Delta M = 0$),
$Y_s(Q^2)$ coincides with the square of the pion form factor, 
\textbf{independently of the energy $\sqrt{s}$ at which it is evaluated},
while
any dependence on $s$ is only due to additional contribution to FSR 
beyond sQED.
In  particular, the difference of $Y_s(Q^2)$ computed at 
two beam energies ($s_1$ and $s_2$), can only come from 
FSR beyond sQED:
\begin{equation}
\Delta Y(Q^2) = Y_{s_1}(Q^2)-Y_{s_2}(Q^2) =  \Delta F_{s_1}(Q^2)- \Delta F_{s_2}(Q^2)
\end{equation}
Therefore, before extracting the pion form factor at threshold, we suggest to 
look at the difference
 $\Delta Y(Q^2)$, which  can be used to estimate the 
contribution beyond sQED to FSR amplitude in a model independent way.


As realistic application of this procedure, 
we consider the case of
DA$\Phi$NE, where KLOE has already 
collected more than 200 pb$^{-1}$ at 1 GeV$^2$ and 
2.5 fb$^{-1}$ at $m_\phi^2$, which, in the range
$Q^2<0.35$ GeV$^2$ , correspond to 
$O(10^3)$ and $O(10^4)$ events respectively in the region
$50^\circ\leq\theta_\gamma\leq 130^\circ$, 
$50^\circ\leq\theta_\pi\leq 130^\circ$. 
We will consider RPT as model for the effects beyond sQED.


\begin{figure}
\label{fig6}
\par
\parbox{1.05\textwidth}{\hspace{-0.3cm}
\includegraphics[width=0.5\textwidth,height=0.5\textwidth]{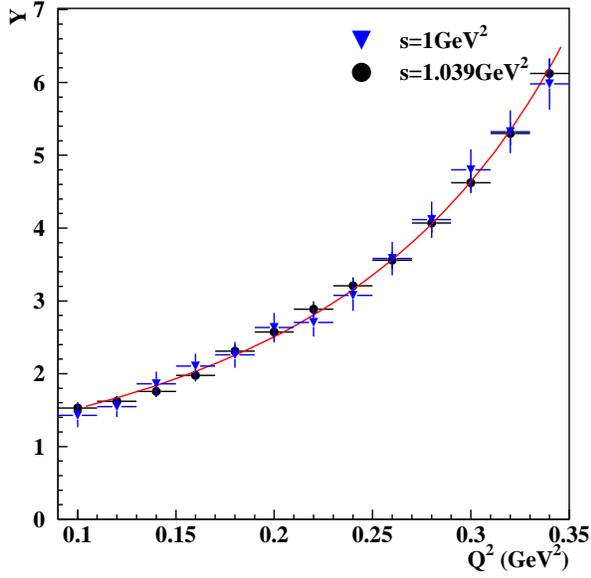}
\hspace{0.3cm}
\includegraphics[width=0.5\textwidth,height=0.5\textwidth]{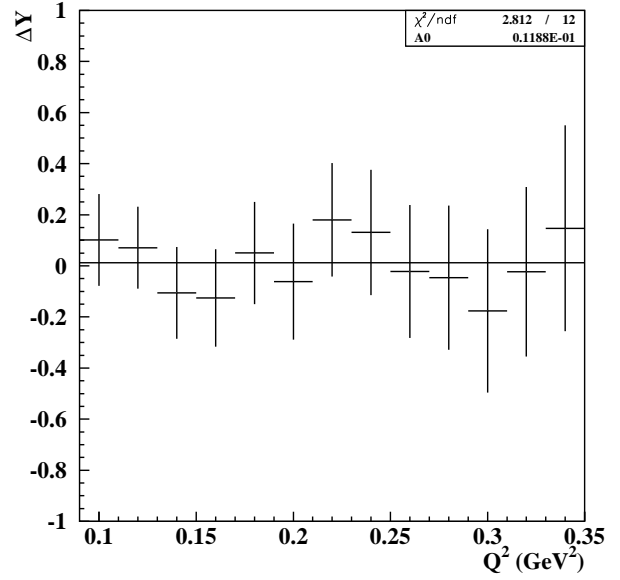}}
\caption{
{\it Left}: $Y_s(Q^2)$ 
at $s=1$ GeV$^2$ (triangles), 
 and at $s=m_\phi^2$  (circles), when FSR includes only sQED and
$\phi$ contribution.  The  pion form factor $|F_{\pi}(Q^2)|^2$ is shown by solid line. 
{\it Right}: The difference $\Delta Y(Q^2)$.
}
\end{figure}
\begin{figure}
\label{fig7}
\par
\parbox{1.05\textwidth}{\hspace{-0.3cm}
\includegraphics[width=0.5\textwidth,height=0.5\textwidth]{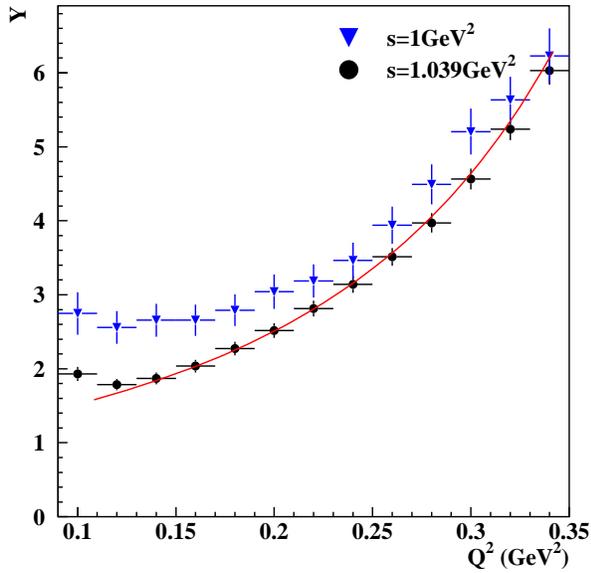}
\hspace{0.3cm}
\includegraphics[width=0.5\textwidth,height=0.5\textwidth]{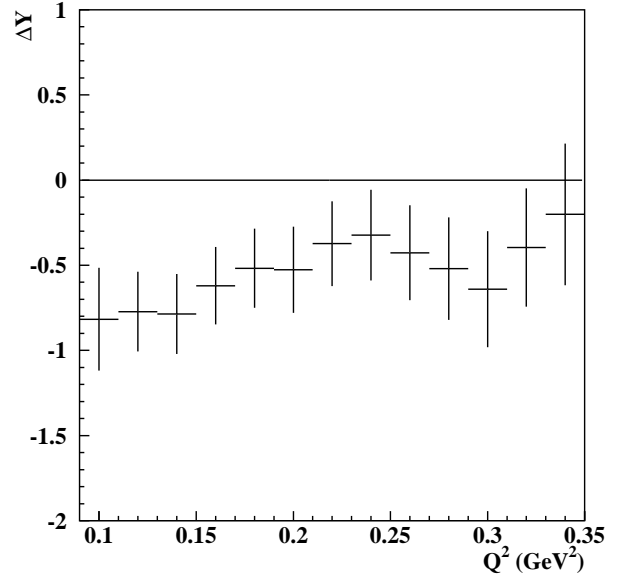}}
\caption{
{\it Left}: $Y_s(Q^2)$  
at $s=1$ GeV$^2$ (triangles), 
 and at $s=m_\phi^2$ (circles), when FSR includes RPT and 
$\phi$ contribution. The
pion form factor $|F_{\pi}(Q^2)|^2$ is shown by solid line. 
{\it Right}: The difference $\Delta Y(Q^2)$.
}
\end{figure}

Fig. 5, {\it left}, shows the quantity $Y_s(Q^2)$ at 
$s_1=1$ GeV$^2$ and at $s_2=m_\phi^2$
 when  no additional RPT term is included in FSR.
As expected, each of these quantities coincides with the square of the 
pion form factor $|F_{\pi}(Q^2)|^2$, shown by solid line. 
The difference $\Delta Y(Q^2)$ is shown in Fig.5,
{\it right}, which is consistent with zero as expected.
A combined fit of $Y_s(Q^2)$ to the pion form factor  (see Eq.~(\ref{decomp_fpi}))
gives the following values:
 $p_1=1.4\pm 0.186$ GeV$^{-2}$, $p_2=8.8\pm 0.73$ GeV$^{-4}$, $\chi^2/\nu=0.25$,
 in agreement with our results at the end of  Sec. 2.2.
 
A different situation appears if FSR emission from pions is modeled by RPT.
In this case, as shown in Fig.6, {\it right}, 
the difference $\Delta Y(Q^2) \neq 0$ and the quantities $Y_s(Q^2)$ 
cannot be anymore identified with $|F_{\pi}(Q^2)|^2$, 
(see Fig.6, {\it left})~\footnote{Destructive interference between RPT and
 $\phi\to\pi^+\pi^-\gamma$ amplitudes tends to cancel out the effects beyond sQED at $s=m_\phi^2$ (see Fig.~2, {\it left}). Therefore the quantity $Y_s(Q^2)$ almost coincides with the pion form factor.}. Prior to the fit of the pion form factor,
in this case, such additional contribution must be understood.


Before concluding, 
we would like to point out the main advantages  of our proposal:

\begin{itemize}
\item NLO correction to ISR (as multi-photon emission) 
can be computed  by Monte Carlo  and included in $H_s$;

\item The quantity  $\Big(\frac{d\sigma_{sQED+\phi}}{dQ^2}\Big)_s$ 
is an input parameter of our procedure, and can be computed numerically
by Monte Carlo;

\item The amplitude for $\phi\to\pi^+\pi^-\gamma$ 
is taken from the  $\pi^0\pi^0\gamma$ channel, therefore 
its description is not restricted to the presence of $f_0$ only,
as in our simulation;

\item A clear advantage of procedure based on 
a Monte Carlo event generator is that it allows to keep control over 
efficiencies and resolutions of the detector and fine tuning of the parameters.

\end{itemize}

Even if  the main limitation of the method could come 
by the uncertainty on the parameters of 
$\phi\to\pi^+\pi^-\gamma$ amplitude, expecially at 
low $Q^2$, we believe 
that the new data  on $\phi\to\pi^0\pi^0\gamma$ 
from KLOE will allow a precise description of this amplitude.
In any case, in agreement with~\cite{czyz} 
we strongly recommend to check the amplitude by using
 charge asymmetry and to compare with
spectrum of the $\pi^+\pi^-\gamma$, at least at high $Q^2$, where the 
pointlike approximation is safe (as done in~\cite{bini}).

\section{Conclusion} 
Test of FSR at threshold  
in the process $e^+e^-\to\pi^+\pi^-\gamma$
 is a rather important issue, not only for the role of FSR as background to 
the measurement of the pion form factor, 
but also to get information about pion-photon interaction when the 
intermediate hadrons are far off shell.
At $s=m_{\phi}^2$ an additional complication arises: the presence of
the direct decay $\phi\to\pi^+\pi^-\gamma$ 
whose amplitude and relative phase can be described according to some model.
By means of a Monte Carlo event generator, which also includes 
the contribution of the direct decay
$\phi\to\pi^+\pi^-\gamma$,
we estimate the effects beyond sQED 
in the framework of Resonance Perturbation Theory (RPT)
for angular cuts used in the
KLOE analysis of the pion form factor at threshold.
We show that the low $Q^2$ region is sensitive 
to the inclusion of additional terms in the FSR amplitude given by the RPT
model.
We propose a method
which allows to estimate the effects beyond sQED in a  model-independent way. We found that the deviation from sQED predicted by RPT can be observed with the current KLOE statistics.





\vspace{1cm}

\textbf{Acknowledgements} 
We thank S. Eidelman,  F. Jegerlehner, H. Czy\'z and K. Melnikov  for useful discussion. We are grateful to W. Kluge and  M. Passera  for careful reading the manuscript and valuable suggestions. G.P. acknowledges  support from EU-CT2002-311 Euridice contract.


\end{document}